\documentstyle[referee]{laa}

\begin{document}
\thesaurus{07  %A&A Section: Solar system
	   (07.09.1; % Interplanetary medium
	   07.13.1;  % Meteoroids
	   07.13.2   % Minor planets
	   )}

%\title{Criteria for Membership in Meteor Streams}
\title{Meteor Stream Membership Criteria}

\author{ Jozef Kla\v{c}ka }
\institute{Institute of Astronomy,
Faculty for Mathematics and Physics,
Comenius University, \\
Mlynsk\'{a} dolina,
842 15 Bratislava,
%Slovak Republic;
%E-mail: klacka@fmph.uniba.sk}
Slovak Republic}
\date{}
\maketitle
\begin{abstract}
Criteria for the membership of individual meteors in meteor streams
are discussed from the point of view of their mathematical and also
physical properties. Discussion is also devoted to the motivation.
It is shown that standardly used criteria
%(mainly D-criterion of Southworth and Hawkins, 1963)
have unusual mathematical properties in the sense of a term ``distance'',
%between points in a phase space, and,
physical motivation and realization for the purpose of obtaining their
final form is not natural and correct, and,
moreover, they lead also to at least surprising astrophysical results.
%General properties of possible criteria are discussed.
A new criterion for the membership in meteor streams is suggested. It
is based on probability theory.
Finally, a problem of meteor
orbit determination for known parent body is discussed.
\end{abstract}

\section{Introduction}
Meteor streams are composed of meteors originating from a parent body
(comet, asteroid) during an ejection process. It is very important to
know the properties of a meteor stream also in order to make some
possible conclusions about physical properties of the parent body,
ejection process(es) and time of possible ejection(s). One of the most
fundamental feature of a meteor stream is the membership of individual
meteors, i. e., if an individual meteor belongs to the given meteor
stream or if it is a background meteor. Meteor stream membership criteria were
suggested for this purpose.

Fundamental characteristics of a body in the Solar System is its orbit
characterized by orbital elements. This holds also for meteors.
Orbital elements of meteors originating from one parent body are similar.
Thus, it is natural that the most simple meteor stream membership criteria
are based on investigation of meteor orbits and their orbital similarity.
Some other criteria may be also used, if it is possible: e. g., physical
composition of individual meteors, based on the observed spectra.

D-criterion of Southworth and Hawkins (1963) is the most frequent
meteor membership criterion used in literature. It measures the orbital
similarity of two individual meteors and thus it enables to find complete
meteor streams. Any meteor is represented by a point in a phase space of
orbital elements. The phase space is a ``five-dimensional orthogonal
coordinate system" (in this section quotation marks are used for the text
taken from Southworth and Hawkins, 1963)
and each element is considered as a coordinate. Southworth and Hawkins take
the following orbital elements: $q$ (perihelion distance), $e$ (eccentricity),
$i$ (inclination), $\Omega$ (longitude of the ascending node), $\omega$
(argument of perihelion). According to the authors, ``the distance between
two points is a natural measure of the difference between the two corresponding
orbits". The orbital similarity function -- ``distance between two points"
(meteors) $A$ and $B$ -- $D(A, B)$ is then defined as
\begin{eqnarray}\label{1}
\left [ D(A, B) \right ] ^{2} &=& \left ( q_{A} ~-~ q_{B} \right ) ^{2} ~+~
	\left ( e_{A} ~-~ e_{B} \right ) ^{2} ~+~
	\left ( 2 ~ \sin \frac{I_{AB}}{2} \right )^{2}	\nonumber \\
& & + ~ \left ( \frac{e_{A} ~+~ e_{B}}{2} \right ) ^{2} ~
	\left ( 2~ \sin \frac{\pi_{AB}}{2} \right ) ^{2} ~,
\end{eqnarray}
where $I_{AB}$ is the angle made by the orbital planes given by the formula
%\begin{eqnarray}\label{2}
%\left ( 2 ~ \sin \frac{I_{AB}}{2} \right )^{2}  &=&
%	 \left ( 2 ~ \sin \frac{i_{A} ~-~i_{B}}{2} \right )^{2}  \nonumber \\
%& & +~  \sin i_{A} ~ \sin i_{B} ~
%	 \left ( 2 ~ \sin \frac{\Omega_{A} ~-~ \Omega_{B}}{2} \right )^{2} ~,
%\end{eqnarray}
\begin{equation}\label{2}
\left ( 2 ~ \sin \frac{I_{AB}}{2} \right )^{2}	=
	\left ( 2 ~ \sin \frac{i_{A} ~-~i_{B}}{2} \right )^{2}
 +~  \sin i_{A} ~ \sin i_{B} ~
	\left ( 2 ~ \sin \frac{\Omega_{A} ~-~ \Omega_{B}}{2} \right )^{2} ~,
\end{equation}
and $\pi_{AB}$ is the difference between the longitudes of perihelion measured
from the point of the orbits intersection
%\begin{eqnarray}\label{3}
%\pi_{AB} &=& \omega_{A} ~-~ \omega_{B} ~+~ 2 ~
%	      \arcsin \left \{ ~ \cos \left ( \frac{i_{A} ~+~ i_{B}}{2} \right ) ~
%	      \times \right . \nonumber \\
%& & \times   \left . \sin \left ( \frac{\Omega_{A} ~-~ \Omega_{B}}{2} \right )
%	      ~\sec \frac{I_{AB}}{2} ~ \right \} ~,
%\end{eqnarray}
\begin{equation}\label{3}
\pi_{AB} = \omega_{A} ~-~ \omega_{B} ~+~ 2 ~
	     \arcsin \left \{ ~ \cos \left ( \frac{i_{A} ~+~ i_{B}}{2} \right ) ~
 \sin \left ( \frac{\Omega_{A} ~-~ \Omega_{B}}{2} \right )
	     ~\sec \frac{I_{AB}}{2} ~ \right \} ~,
\end{equation}
or, to a sufficient approximation,when $i_{A}$ and $i_{B}$ are small,
\begin{equation}\label{4}
\pi_{AB} = \left ( \Omega_{A} ~+~ \omega_{A} \right ) ~-~
	     \left ( \Omega_{B} ~+~ \omega_{B} \right ) ~.
\end{equation}
This D-criterion is used for identification of meteor streams: if
$D(A, B) < D_{c}$, where $D_{c}$ is a constant assumed as a threshold value,
then orbits of meteors $A$ and $B$ are similar and both meteors may be members
of the same meteor stream.

A little modified D-criteria were suggested by Drummond (1981) and Jopek (1993).

The aim of this article is to analyze mathematical and physical properties
of Southworth-Hawkins' D-criterion (properties of Drummond's and Jopek's
D-criteria are analogous). Finally, we present a new meteor stream
membership criterion.

\section{Mathematics and D-criterion}
In the construction of their D-criterion, Southworth and Hawkins were originally
inspired by the well-known definition of the distance in Euclidean space.
As a consequence they have obtained D-criterion in the form (1) (and
also Eqs. (2)-(4)). As it is considered, D-criterion measures the distance
between orbits of two meteoroids (meteors). However, if it is so, then
D-criterion (1) must fulfill properties required for a quantity
called distance. The standard properties of a distance are closely connected
with the so-called metric space. Definition states: \\

Let $X$ be a set with elements $u, v, w, ...$. A nonnegative function $\rho$
defined on the Cartesian product $X \times X$ is called a {\it metric} if it
satisfies the following axioms: \\
\hspace*{0.1cm}   (i) ~~~$\rho (u, v) =$ ~0 ~ if and only if ~ $u = v$ ~;  \\
\hspace*{0.00cm} (ii) ~~~$\rho (u, v) = ~\rho (v, u)$ ~; \\\
(iii) ~~~$\rho (u, v) \le ~ \rho (u, w) ~+~ \rho (w, v)$ ~. \\
A set $X$ with a metric $\rho$ is called a {\it metric space}. \\
(Metric -- distance. The property (iii) is called {\it the triangle inequality}.) \\

Now, question is, if these properties are fulfilled also for D-criterion (1).
One can easily verify that triangle inequality is violated. It means that
triangle inequality, which is an evident property of a distance, is not
fulfilled in the case of measuring ``distances'' between meteor (meteoroid)
orbits.

As an evident property of the D-criterion we introduce the following one. If
$D(u, v)$ is smaller than $D(u, w)$, then orbits $u$ and $v$ are more
similar than the orbits $u$ and $w$. However, due to the violation of the
triangle inequality, the orbits $v$ and $w$ may be more similar than one would
expect on the basis of general conception about distance: \\

0 ~$< ~D(v, w)~ < ~D(u, w) ~-~ D(u, v)$ ~.\\

We can formulate this in terms of meteor orbits: the distance between meteors
$u$ and $v$ is small, the distance between meteors $u$ and $w$ is large, but
the distance between meteors $v$ and $w$ may be small.

From the mathematical point of view it would be useful to have a D-criterion
which fulfills the properties (i)-(iii). It may have also other forms than
presented by Eq. (1), i. e., it may contain not only terms of the form
$( \alpha_{j} ~-~ \alpha_{k} ) ^{2}$, but also, e. g., terms of the form
$| \alpha_{j} ~-~ \alpha_{k} |$, and so on. We know from mathematics that
the number of possible (from a mathematical point of view) metrics is infinity:
if $\rho (x, y)$ is a metric, then
$\rho_{1} (x, y) = \rho (x, y) / \left \{ 1 ~+~ \rho (x, y) \right \} $
is also a metric. One of the most simple modifications of Eq. (1) satisfying
properties (i)-(iii) may be obtained by substituting the term
$\left ( e_{A} ~+~ e_{B} \right ) / 2$ by any constant; the consequence of
the modification
$\left ( e_{A} ~+~ e_{B} \right ) / 2~ \longrightarrow ~1$
is, e. g., that the group of Cylids
(Southworth and Hawkins, 1963, p. 271) consists of two meteor streams
($D_{SH} =$ 0.06, 0.10, 0.10, 0.10, 0.14; $D_{new} =$ 0.24, 0.19, 0.59,
0.59, 0.77).

As for meteors, standard procedure is to choose one orbit as a reference orbit.
The reference orbit may be given by mean values of the known members of a given
meteor stream. Any meteor $A$ is a member of the stream if $D(A, M) < D_{c}$,
where $M$ represents an orbit defined by mean values of orbital
elements and $D_{c}$ is a constant assumed to be a threshold value. Since
$D(A, M)$ is given by Eqs. (1) -- (4), we may have $D(A, M) < D_{c}$ and
$D(B, M) < D_{c}$ for some two meteors $A$ and $B$ of the stream, but
$D(A, B) > 2~D_{c}$. In other words: the first two inequalities assert that
meteors $A$ and $B$ have similar orbits but the last inequality states that
their orbits are not similar! (Points $A$ and $B$ are situated inside a sphere
with a centre $M$ and a radius $r$, but the distance between $A$ and $B$ may be
greater than 2$r$!) This is the consequence of the triangle inequality
violation. (If $D \rightarrow$ 0 and, moreover, $\Delta e / e \ll$ 1,
$\sin \Delta \pi / 2 \ll$ 1 (i. e., orbits are identical or almost
identical), then the violation of the triangle inequality plays no role.
However, this may not be the case occuring in astronomical applications
(see Kla\v{c}ka and Volo\v{s}in 1996).)

Those, who are interested in mathematical properties of semi-metric
(or, even in a more general case of semi-pseudometric) defined by Eq. (1),
we refer to section 18 in \v{C}ech (1966).

\section{Physics and D-criterion}

Southworth and Hawkins (1963) present also physical arguments for the choice
of D-criterion in the form of Eqs. (1)-(4). However, their arguments are not
convincing and thus one should take Eq. (1) as an empirical criterion.
We will justify this statement now.

The physical model of Southworth and Hawkins is based on the idea that the
change of meteoroid's orbital elements with respect to those of the parent body
may be represented as an average value of the changes (weighted by velocity)
during one period of the parent body. However, this idea has no advantage.
Nor it is a simple idea (hypothesis), nor does correspond to real processes:
the meteoroid is ejected at once (at one moment) from the parent body.

Moreover, mathematical realization of the physical model is incorrect and some
other physical inconsistencies arise in the process of mathematical calculations.
According to the idea, we should write for the change of any orbital element $G$

\noindent
$\Delta G = \left ( \nabla _{\vec{v}} G \right ) \cdot \left ( \Delta \vec{v} \right )$ ~.

\noindent
However, the authors use in their calculations

\noindent
$\Delta G = | \nabla _{\vec{v}} G | ~ | \Delta \vec{v} | $~,

\noindent
which is different from the correct value, since, in general, $\Delta \vec{v}$
is not in the direction of $\nabla _{\vec{v}} G$:

\noindent
%{\Large
%$ - 1 \leq \frac{\left ( \nabla _{\vec{v}} G \right ) \cdot \left ( \Delta \vec{v} \right )}{
%	   | \nabla _{\vec{v}} G | ~ | \Delta \vec{v} |} \leq + 1 $~.} \\

$ - 1 \leq \left \{ \left ( \nabla _{\vec{v}} G \right ) \cdot \left ( \Delta \vec{v} \right )
	   \right \} / \left \{
	   | \nabla _{\vec{v}} G | ~ | \Delta \vec{v} | \right \} \leq + 1 $~. \\

The consequence of these three equations is that even if the model of the
authors would be correct, its realization is incorrect -- they add together not
changes of orbital elements, but their extremal values with positive
signs at each instant. Although this would seem acceptable in the sense of
calculations of maximum possible changes -- which is not the author's interpretation,
and, even if it would be --, it is incorrect: for the same meteoroid,
$\Delta \vec{v}_{R}$ may be dominant for one orbital element,
$\Delta \vec{v}_{S}$ may be dominant for another orbital element
(see Eqs. (A8) in Southworth and Hawkins (1963); we used the same notation
as Southworth and Hawkins). The other important nonphysical result rests
in the assumption that $| \Delta \vec{v}|$ is proportional to circular
velocity at the distance $r$:

\noindent
$ | \Delta \vec{v} | \propto U~, ~i. e.,~~~ | \Delta \vec{v} | \propto r^{-1/2}$~.

\noindent
(In the appendix of Southworth and Hawkins, there should be

\noindent
$\Delta G \propto \int r^{-1/2} ~| \nabla _{\vec{v}} G | ~dt ~$ .

\noindent
Fortunately, calculations of the authors correspond to this result.)

I am not sure about the correctness of the final results of
Southworth and Hawkins (their appendix and Fig. 1), e. g., my result is:
$\lim _{e \rightarrow 0} \left \{ \left ( 1 / a \right ) ~Grad ~q \right \}
= 0.64$,
%$\lim _{e \rightarrow 0} Grad \pi = \lim _{e \rightarrow 0} Grad e = 1.54$,
which seems to be not consistent with results depicted on Fig. 1 in
Southworth and Hawkins' article. (The mean values calculated in appendix
of the authors are calculated from absolute values of quantities, so as
the mathematical lemma could be used.)

According to Fig. 1 in
Southworth and Hawkins (1963), the authors make a conclusion that their
D-criterion may be applied for $e < 0.85$. In reality, it is used even for
$e \approx$ 1. Individual terms of the sum in Eq. (1) are comparable in their
values also for $e >$ 0.85, which is not consistent with Fig. 1 in
Southworth and Hawkins. This also shows that Eq. (1) is not consistent with
the physics suggested by the authors.

\section{Astronomy and D-criteria}

On the basis of the previous two sections we can conclude that meteor stream
membership criteria
(Southworth and Hawkins 1963, Drummond 1981, Jopek 1993) have two
unpleasant properties, as for mathematics and physics: \\
i) triangle inequality does not hold for arbitrary orbits, \\
ii) physics of the criteria is unknown.

If we want to make a simple physical model, we may imagine that meteoroids
are ejected only at parent body's perihelion and calculate the change of
meteoroid's orbital elements with respect to those of the parent body. One
can obtain
\begin{eqnarray}\label{5}
\left ( \frac{\Delta \vec{v}}{v_{PB}} \right ) ^{2} &=&
 \frac{ \left ( p_{A} ~-~ p_{M} \right ) ^{2}}{8~ p_{M}^{2}} ~+~
 \frac{ \left ( e_{A} ~-~ e_{M} \right ) ^{2}}{8~ \left ( 1 ~+~ e_{M} \right )^{2}}
 ~+ \nonumber \\
& &	\left ( \sin I_{AM} \right )^{2}  ~+~
      ~ \left ( \frac{e_{M}}{1 ~+~ e_{M}} \right ) ^{2} ~
	\left ( \sin \pi_{AM} \right ) ^{2} ~,
\end{eqnarray}
where $v_{PB}$ is parent body's speed at perihelion (ejection of the meteoroid $A$),
subscript $M$ corresponds to the parent body; $p = a ( 1~-~e )$, $a$ -- semimajor axis.
We may define, on the basis of Eq. (5) the following quantity:
%(symmetricity seems to be violated if one of $A$ or $B$ equals $M$,
%but multiplication terms are numerical constants for a given meteor stream):
%\begin{eqnarray}\label{6}
%\left [ D(A, B) \right ] ^{2} &=&
% \frac{ \left ( p_{A} ~-~ p_{B} \right ) ^{2}}{8~ p_{M}^{2}} ~+~
% \frac{ \left ( e_{A} ~-~ e_{B} \right ) ^{2}}{8~ \left ( 1 ~+~ e_{M} \right )^{2}}
%  ~+ \nonumber \\
%& &	 \left ( \sin I_{AB} \right )^{2}  ~+~
%      ~ \left ( \frac{e_{M}}{1 ~+~ e_{M}} \right ) ^{2} ~
%	 \left ( \sin \pi_{AB} \right ) ^{2} ~.
%\end{eqnarray}
\begin{eqnarray}\label{6}
\left [ D(A, M) \right ] ^{2} &=& \alpha ~
				  \left ( p_{A} ~-~ p_{M} \right ) ^{2} ~+~
				  \beta ~ \left ( e_{A} ~-~ e_{M} \right ) ^{2}
				  ~+ \nonumber \\
& &	\left ( \sin I_{AM} \right )^{2}  ~+~ \gamma ~
	\left ( \sin \pi_{AM} \right ) ^{2} ~,
\end{eqnarray}
where $\alpha$, $\beta$ and $\gamma$ are numerical constants
for a given meteor stream: $\alpha = 1 / ( 8~ p_{M}^{2} )$,
$\beta = 1 /  ( 8~ ( 1 ~+~ e_{M} )^{2} )$,
$\gamma = ~ e_{M}^{2} / ( 1 ~+~ e_{M} ) ^{2}$.
The first two terms of the sum are equal in the problem of two bodies.
In reality, however,
they are often not comparable (some terms of the sum are often negligible
in comparison with the others) for real meteor streams. This shows that although
physics of Eq. (6) is simple, it does not correspond to useful meteor stream
membership criterion.

Practical advantage of the Southworth and Hawkins' criterion in comparison
with the criterion defined in Eq. (6) is that the individual terms of the sum
are more comparable than it is in the case of Eq. (6). However, if this is the
only requirement for the choice of the criterion for practical usage, then
we can write better criterion at once (numerical factor $\xi = e_{M}$):
%\begin{eqnarray}\label{7}
%D(A, B) &=& | q_{A} ~-~ q_{B} | ~+~| e_{A} ~-~ e_{B} | ~+~
%	 2 ~ |	\sin \frac{I_{AB}}{2} |  ~+ \nonumber \\
%& &	 2 ~ \xi ~| \sin \frac{\pi_{AB}}{2} | ~.
%\end{eqnarray}
\begin{equation}\label{7}
D(A, B) = | q_{A} ~-~ q_{B} | ~+~| e_{A} ~-~ e_{B} | ~+~
	2 ~ |  \sin \frac{i_{A} ~-~ i_{B}}{2} |  ~+ ~
     2 ~ \xi ~| \sin \frac{\pi_{A} ~-~\pi_{B}}{2} | ~.
\end{equation}
In practice, the quantity $e_{M}$ should be calculated as the mean value of
eccentricities of bodies forming given meteor stream.
Individual terms of the sum of Eq. (7) are more comparable than it is in Eq. (1)
(moreover, triangle inequality is also fulfilled;
remark: Eq. (6) does not represent any distance -- the index M cannot be
chaanged into B !).
Many modifications of
Eq. (7) may be used, e. g., $|q_{A} - q_{B}| \rightarrow |q_{A} - q_{B}| /
(q_{A} + q_{B})$, the same substitution for eccentricities, their various
combinations, etc..

There is no reason for the fact that individual terms of the
sum in Eqs. (1) or (7) should be statistically comparable for a
given meteor stream. On the contrary, one should expect that
some terms are dominant for some meteor streams, other terms may
be more important for other meteor streams. {\it Terms of the sum should be
weighted.}

There are other unpleasant properties of the membership criteria of
Southworth and Hawkins (1963), Drummond (1981), Jopek (1993). All of them use
perihelion distance $q$ as an orbital element. The consequence is that
%even meteors with semi-major axis $a$ greater than 10 AU (or even 100 AU)
meteors with large dispersions in semi-major axis $a$
may be classified as members of a given stream, which does not seem to be real
-- there is very small probability that one or several meteoroids can be
ejected with velocities much greater (even in orders of magnitude) than
the other meteoroids from the same parent body (or bodies of comparable
properties). But, when we would like to make a substitution $q \rightarrow a$,
or, $p \rightarrow a$ (e. g., in Eqs. (1) and (7)) the individual terms in
$D(A, B)$ are much less comparable and usually the term with $a$ is dominant
(probably, weighting coefficients might solve this situation).
The last remark: many people using D-criteria incorrectly calculate the mean
values of angular quantities.

In conclusion of this section we can make a statement that it is not possible
to give a simple D-criterion based on physical arguments. Thus, one can write
many forms of D-criteria. If we would apply various D-criteria on meteor streams
(and their background) containing several tens of meteors, we should expect more
than 80 \% (or, perhaps, even more than 90 \%) coincidence between them.
If we would compare them with the results obtained by Tisserand parameter,
the coincidence may even decrease to about 50 -- 70 \%. (Of course, it has
no sense to use various D-criteria for meteor streams containing less than
10 known members -- various D-criteria lead to completely different results.)

All statements made in the last paragraph should be understand in the way that
it is not very useful to use D-criteria based only on the problem of two bodies.
Tisserand parameter, invariant of the motion in the restricted three-body problem,
may be a better criterion. However, one must bear in mind that meteoroids are perturbed
on their orbits not only by gravitational forces. Nongravitational forces may
also be very important. And some of them may be of stochastic nature. This is
the reason of our suggestion presented in the following section.

\section{Probability, Statistical Mathematics and D-criteria}

Since we do not know any simple physics which can define in a simple way a
meteor stream, we take the data (set of orbital elements for various
meteors) as a random sample.

Since four orbital elements completely define a meteor stream (intersecting
the orbit of the Earth), we will take four orbital elements $Q_{1}$,
$Q_{2}$, $Q_{3}$, $Q_{4}$. Let the real distribution of orbital elements
of meteors in the stream may be approximated by density function
$f$ ($Q_{1}$, $Q_{2}$, $Q_{3}$, $Q_{4}$) $\equiv$ $f (X)$. If we define
the meteor stream as a set of bodies with $X \in \Omega$,
\begin{equation}\label{8}
P ( X \in \Omega ) \equiv \int_{\Omega} f ( X ) d X = \alpha~,
\end{equation}
then there is a probability less than 1 ~--~ $\alpha$ that objects with
$X \in \Omega '$ belong to the stream.

\section{Meteor Orbit Determination}

Another problem concerning the orbits of meteors is the problem of
finding radiants for given parent bodies.
In determining orbital elements for a meteor corresponding to the meteoroid $A$
initially ejected from some parent body $B$, we must use the following sets of
equations:
%\begin{eqnarray}\label{9}
%\frac{\partial F}{\partial \beta_{iA}} ~+~
%\lambda~ \frac{\partial }{\partial \beta_{iA}}
%\left \{ \frac{p_{A}}{1 ~+~ e_{A}~ \cos \omega_{A}} \right \}
%&=& 0 ~,~ i = 1 ~to~ 5 ~, \nonumber \\
%\frac{p_{A}}{1 ~+~ e_{A}~ \cos \omega_{A}} &=& 1 ~,
%\end{eqnarray}
\begin{equation}\label{9}
\frac{\partial F}{\partial \beta_{iA}} ~+~
\lambda~ \frac{\partial }{\partial \beta_{iA}}
\left \{ \frac{p_{A}}{1 ~+~ e_{A}~ \cos \omega_{A}} \right \}
= 0 ~,~ i = 1 ~to~ 5 ~,~~~
\frac{p_{A}}{1 ~+~ e_{A}~ \cos \omega_{A}} = 1 ~,
\end{equation}
corresponding to the case if the ascending node intersects the orbit of the
Earth, and
%\begin{eqnarray}\label{10}
%\frac{\partial F}{\partial \beta_{iA}} ~+~
%\lambda~ \frac{\partial }{\partial \beta_{iA}}
%\left \{ \frac{p_{A}}{1 ~-~ e_{A}~ \cos \omega_{A}} \right \}
%&=& 0 ~,~ i = 1 ~to~ 5 ~, \nonumber \\
%\frac{p_{A}}{1 ~-~ e_{A}~ \cos \omega_{A}} &=& 1 ~,
%\end{eqnarray}
\begin{equation}\label{10}
\frac{\partial F}{\partial \beta_{iA}} ~+~
\lambda~ \frac{\partial }{\partial \beta_{iA}}
\left \{ \frac{p_{A}}{1 ~-~ e_{A}~ \cos \omega_{A}} \right \}
= 0 ~,~ i = 1 ~to~ 5 ~, ~~~
\frac{p_{A}}{1 ~-~ e_{A}~ \cos \omega_{A}} = 1 ~,
\end{equation}
for the case if the descending node intersects the Earth's orbit.
$\beta_{iA}$ correspond to orbital elements
of the meteoroid (body $A$), the constraint
corresponds to the fact that
meteoroid may strike the Earth (circular orbit of the Earth is supposed, for the
sake of simplicity).
The quantity $F$ may be of the type
\begin{eqnarray}\label{11}
F &=& \alpha_{1} ~ ( E_{A} ~-~ E_{B} )^{2} ~+~
      \alpha_{2} ~ ( \vec{L_{A}} ~-~ \vec{L_{B}} )^{2} ~+~ \nonumber \\
& &   \alpha_{3} ~ ( \vec{v_{TA}} ~-~ \vec{v_{TB}} )^{2} ~+~
      \alpha_{4} ~ ( \vec{v_{RA}} ~-~ \vec{v_{RB}} )^{2} ~ +~ \nonumber \\
& &   \alpha_{5} ~ ( \vec{v_{TfA}} ~-~ \vec{v_{TfB}} )^{2} ~,
\end{eqnarray}
where
\begin{eqnarray}\label{12}
( E_{A} ~-~ E_{B} )^{2} &=& \left ( \frac{1 ~-~ e_{A}^{2}}{p_{A}} ~-~
			    \frac{1 ~-~ e_{B}^{2}}{p_{B}} \right ) ^{2} ~,
			    \nonumber \\
( \vec{L_{A}} ~-~ \vec{L_{B}} )^{2} &=& p_{A} ~+~ p_{B}~ ~-~
			    2~ \sqrt{p_{A}~p_{B}} ~ \times \nonumber \\
& &			    \left \{
			    \cos \left ( \Omega_{A} ~-~ \Omega_{B} \right ) ~
%			     \sin i_{A} ~ \sin i_{B} ~+~ \right . \nonumber \\
%& &  \left .		     \cos i_{A} ~ \cos i_{B} ~
%			     \right \} ~, \nonumber \\
			    \sin i_{A} ~ \sin i_{B} ~+~
			    \cos i_{A} ~ \cos i_{B} ~
			    \right \} ~, \nonumber \\
( \vec{v_{TA}} ~-~ \vec{v_{TB}} )^{2} &=&
       \frac{\left ( 1 ~+~ e_{A} \right )^{2}}{p_{A}} ~+~
       \frac{\left ( 1 ~+~ e_{B} \right )^{2}}{p_{B}} ~-~ \nonumber \\
& &    2~ \frac{\left ( 1 ~+~ e_{A} \right ) \left ( 1 ~+~ e_{B} \right )~}
       {\sqrt{p_{A}~p_{B}}} \times X ~, \nonumber \\
X&=& \cos (\Omega_{B} ~-~ \Omega_{A}) ~ \{ \sin \omega_{A} ~ \sin \omega_{B} ~+~
     \nonumber \\
& &  \cos \omega_{A} ~ \cos \omega_{B} ~ \cos i_{A} ~ \cos i_{B} \} ~+~ \nonumber \\
& &  \sin (\Omega_{B} ~-~ \Omega_{A}) \{ \sin \omega_{A} ~ \cos \omega_{B} ~
     \cos i_{B} ~-~ \nonumber \\
& &  \sin \omega_{B} ~ \cos \omega_{A} ~ \cos i_{A} \} ~+~ \nonumber \\
& &  \cos \omega_{A} ~\cos \omega_{B} ~ \sin i_{A} ~ \sin i_{B} ~, \nonumber \\
( \vec{v_{RA}} ~-~ \vec{v_{RB}} )^{2} &=&
       \frac{e_{A}^{2}}{p_{A}} ~+~
       \frac{e_{B}^{2}}{p_{B}} ~-~
       2~ \frac{e_{A} ~ e_{B}}
       {\sqrt{p_{A}~p_{B}}} \times Y ~, \nonumber \\
Y&=& \cos (\Omega_{B} ~-~ \Omega_{A}) ~ \{ \cos \omega_{A} ~ \cos \omega_{B} ~+~
     \nonumber \\
& &  \sin \omega_{A} ~ \sin \omega_{B} ~ \cos i_{A} ~ \cos i_{B} \} ~+~ \nonumber \\
& &  \sin (\Omega_{B} ~-~ \Omega_{A}) \{ \sin \omega_{A} ~ \cos \omega_{B} ~
     \cos i_{A} ~-~ \nonumber \\
& &  \sin \omega_{B} ~ \cos \omega_{A} ~ \cos i_{B} \} ~+~ \nonumber \\
& &  \sin \omega_{A} ~\sin \omega_{B} ~ \sin i_{A} ~ \sin i_{B}~, \nonumber \\
( \vec{v_{TfA}} ~-~ \vec{v_{TfB}} )^{2} &=&
       \frac{\left ( 1 ~-~ e_{A} ~/~ \sqrt{2} \right )^{2}}{p_{A}} ~+~ \nonumber \\
& &    \frac{\left ( 1 ~-~ e_{B} ~/~ \sqrt{2} \right )^{2}}{p_{B}} ~-~ \nonumber \\
& &    \frac{\left ( 1 ~-~ e_{A} ~/~ \sqrt{2} \right ) \left ( 1 ~-~ e_{B} ~/~ \sqrt{2} \right )~}
       {\sqrt{p_{A}~p_{B}}} \times Z ~, \nonumber \\
Z&=& \cos (\Omega_{B} ~-~ \Omega_{A}) \times~ \nonumber \\
& & \{
     ( \sin \omega_{A} ~-~ \cos \omega_{A} ) ~ ( \sin \omega_{B} ~-~ \cos \omega_{B} )	~+~
     \nonumber \\
& &  ( \sin \omega_{A} ~+~ \cos \omega_{A} ) ~ ( \sin \omega_{B} ~+~ \cos \omega_{B} ) \times \nonumber \\
& &  \cos i_{A} ~ \cos i_{B} \} ~+~ \nonumber \\
& &  \sin (\Omega_{B} ~-~ \Omega_{A}) ~ \times ~ \nonumber \\
& &  \{ ( \sin \omega_{A} ~-~ \cos \omega_{A} ) ~
     ( \sin \omega_{B} ~+~ \cos \omega_{B} ) ~ \times \nonumber \\
& &  \cos i_{B} ~-~ \nonumber \\
& &  ( \sin \omega_{A} ~+~ \cos \omega_{A} ) ~ ( \sin \omega_{B} ~-~ \cos \omega_{B} )
     ~ \times \nonumber \\
& &  ~ \cos i_{A} \} ~+~ \nonumber \\
& &  ( \sin \omega_{A} ~+~ \cos \omega_{A} ) ~ ( \sin \omega_{B} ~+~ \cos \omega_{B} ) \times \nonumber \\
& &   ~ \sin i_{A} ~ \sin i_{B} ~.
\end{eqnarray}
One of the coefficients $\alpha_{1} -  \alpha_{5}$ may be put equal to 1
(e. g., $\alpha_{1} =$ 1). $\alpha-$coefficients
are functions of orbital elements
and their values must be determined from known pairs ``parent body -- meteor
stream''. As for the physical sense of the individual terms of the sum in
Eq. (11), they correspond to: energy, angular momentum, perihelion velocity
and radial velocity of maximum value, transversal velocity for true anomaly
$f = 3~ \pi ~/ ~4$ in the problem
of two bodies. The $\alpha$-coefficients guarantee that Eqs. (9)-(12) can be
applied on real situations. The advantage of the form of Eq. (11) is that it
contains known physical quantities. In principle, we can choose other forms.
However, the requirement that they must contain all five orbital elements
in an independent way may not suffice in obtaining good coincidence
between theoretical and observed radiants.
Local and global minima may be important, in general.

%Perhaps, knowing $\alpha$-coefficients, Eqs. (14)-(15) may also be used
%as a meteor stream membership criterion.

\section{Conclusion}
We have shown that standardly used method for determining meteor stream
membership is not correct from the point of view of mathematics, physics
and astronomy. We have presented correct method, based on probability
theory and statistical mathematics.

\acknowledgements
Special thanks to the firm ``Pr\'{\i}strojov\'{a} technika, spol. s r. o.''.
This work was partially supported by Grants VEGA No. 1/4304/97 and
1/4303/97. The author wants to thank to the organising committee of the
ACM'96 for the possibility of presenting this paper at the conference.
Small comments of referees of both papers (the second paper -- Meteor
streams and parent bodies) are also acknowledged, especially that of
T. Jopek which enabled to remove one incorrect sign (1995).

%This work was supported by Grant No.~B-02-028 -- ESO C\&EE Programme. The author
%wants to thank to D. Steel for the careful reading of the paper, critical
%comments and some other suggestions aiming at improving the quality of the
%paper. Also comments of the referee are acknowledged.


\begin{thebibliography}{}
%\bibitem{}Asher, D. J., Clube, S. V. M., Steel, D. I., 1993, in: Meteoroids
%and Their Parent Bodies, eds. J. \v{S}tohl and I. P. Williams,
%Slovak Academy of Sciences, Bratislava, p. 93
%\bibitem{}Brouwer, D., van Woerkom, A. J. J., 1950, Astron. Papers Amer.
%Ephemeris 13, 83
\bibitem{}\v{C}ech, E., 1966, Topological Spaces, Academia, Prague, 893 pp.
\bibitem{}Drummond, J. D., 1981, Icarus 45, 545
\bibitem{}Jopek, T. J., 1993, Icarus 106, 603
\bibitem{}Kla\v{c}ka, J., Volo\v{s}in, A., 1996 (presented at ACM'96)
%\bibitem{}Kla\v{c}ka, J., Volo\v{s}in, A., 1996 (to be submitted to
%Astronomy and Astrophysics)
%\bibitem{}Kla\v{c}ka J., 1987, Thesis, Comenius University, Bratislava
%(in Slovak)
%\bibitem{}Kla\v{c}ka J., 1991, Earth, Moon and Planets 55, 45
%\bibitem{}Kla\v{c}ka J., 1994, Earth, Moon and Planets (in press)
%\bibitem{}Kla\v{c}ka J., Pittich, E. M., 1994, presented at the conference
%{\it Meteoroids}, Bratislava (to be published)

%\bibitem{}Dohnanyi J. S., 1978, in: Cosmic Dust, ed. J. A. M. McDonnell, Wiley,
%Chichester, p. 527
%\bibitem{}Robertson H. P., 1937, MNRAS 97, 423
%\bibitem{}Robertson H. P., Noonan, T. W., 1968, Relativity and Cosmology,
%W. B. Saunders Company, Philadelphia
\bibitem{}Southworth, R. B., Hawkins, G. S., 1963, Smithson. Contrib. Astrophys.
7, 261
\end{thebibliography}
\end{document}